\documentclass[preprint]{elsarticle}
\usepackage[utf8]{inputenc}
\usepackage[T1]{fontenc}
\usepackage[bottom]{footmisc}
\usepackage{hyphenat}
\usepackage{times, amsfonts, amssymb, amsmath, textcomp, gensymb, epsfig, makeidx, graphicx, graphics, url, multirow, multicol, booktabs, array, bm, rotating, float, setspace, paralist, subfig, wrapfig, color, enumitem, latexsym, longtable, tabularx, xcolor, mathtools, stmaryrd, commath, tikz, soul, lscape, svg, natbib} %endfloat
\usepackage{underscore}
\DeclareMathAlphabet{\mathcal}{OMS}{cmsy}{m}{n}

\graphicspath{{figures/}}

\begin{document}

\begin{frontmatter}
 
\title{\texttt{instancespace}: a Python Package for Insightful Algorithm Testing through Instance Space Analysis}

\author[aff:comp]{Yusuf Berdan Güzel}
\ead{y.guzel@unimelb.edu.au}
\author[aff:comp]{Kushagra Khare}
\ead{kharek@student.unimelb.edu.au}
\author[aff:comp]{Nathan Harvey}
\author[aff:comp]{Kian Dsouza}
\author[aff:comp]{Dong Hyeog Jang}
\author[aff:comp]{Junheng Chen}
\author[aff:comp]{Cheng Ze Lam}
\author[aff:comp,aff:optima]{Mario Andrés Muñoz\corref{cor1}}
\ead{munoz.m@unimelb.edu.au}
\cortext[cor1]{Corresponding author}

\affiliation[aff:comp]{organization={School of Computing and Information Systems, The University of Melbourne},
            addressline={Parkville}, 
            city={Melbourne},
            postcode={3010}, 
            state={VIC},
            country={Australia}}

\affiliation[aff:optima]{organization={ARC Training Centre in Optimisation Technologies, Integrated Methodologies, and Applications (OPTIMA)},
            addressline={Parkville}, 
            city={Melbourne},
            postcode={3010}, 
            state={VIC},
            country={Australia}}

\begin{abstract}
    Instance Space Analysis is a methodology to evaluate algorithm performance across diverse problem fields. Through visualisation and exploratory data analysis techniques, Instance Space Analysis offers objective, data-driven insights into the diversity of test instances, algorithm behaviour, and algorithm strengths and weaknesses. As such, it supports automated algorithm selection and synthetic test instance generation, increasing testing reliability in optimisation, machine learning, and scheduling fields. This paper introduces \texttt{instancespace}, a Python package that implements an automated pipeline for Instance Space Analysis. This package supports research by streamlining the testing process, providing unbiased metrics, and facilitating more informed algorithmic design and deployment decisions, particularly for complex and safety-critical systems.
\end{abstract}


\end{frontmatter}

% \section*{Metadata}

\begin{table}[!t]
\centering
\begin{tabular}{|l|p{5.0cm}|p{5.0cm}|}
\hline
\textbf{Nr.} & \textbf{Code metadata description} & \textbf{Please fill in this column} \\
\hline
C1 & Current code version & v0.2.1\\
\hline
C2 & Permanent link to code/repository used for this code version & \url{https://github.com/andremun/pyInstanceSpace}\\
\hline
C3  & Permanent link to Reproducible Capsule &  \url{https://github.com/andremun/pyInstanceSpace/blob/main/instancespace/example.ipynb}\\
\hline
C4 & Legal Code License   & GPL-3.0\\
\hline
C5 & Code versioning system used & Git\\
\hline
C6 & Software code languages, tools, and services used & Python \\
\hline
C7 & Compilation requirements, operating environments \& dependencies & 

python = "+3.12"

numpy = "+1.26.4"

pandas = "+2.2.1"

pandas-stubs = "+2.2.1.240316"

click = "+8.1.7"

scipy = "+1.13.0"

pygad = "+3.3.1"

shapely = "+2.0.5"

matplotlib = "+3.9.2"

alphashape = "+1.3.1"

loguru = "+0.7.2"

scikit-optimize = "+0.10.2"

scikit-learn = "+1.5.2"
\\
\hline
C8 & Link to developer documentation &  \url{https://github.com/andremun/pyInstanceSpace/blob/main/README.md} \\
\hline
C9 & Support email for questions & 
munoz.m@unimelb.edu.au
y.guzel@unimelb.edu.au
\\
\hline
\end{tabular}
\caption{Code metadata}
\label{codeMetadata} 
\end{table}

\section{Motivation and Significance}

A challenge in algorithm testing, particularly in Artificial Intelligence-related fields such as Optimisation, Machine Learning and Time Series Analysis, is using benchmarking test instances that fail to meet essential diversity criteria. To reveal the strengths and weaknesses of algorithms, a suite of test instances should not only be unbiased and challenging. It should also include a mix of synthetically generated and real-world-like instances with sufficiently diverse structural properties. Since the conclusions drawn during benchmarking critically depend on the choice of test instances~\cite{Hooker95}, without such diversity, the reliability of the analysis is compromised. However, standard practice involves testing algorithms in well-known test suites without necessarily examining their suitability. Meanwhile, substantial focus is placed on reporting average performance across the suite, disregarding potential weaknesses or how performance depends on the properties of the test instances. 

Instance Space Analysis (ISA) is a methodology designed to address these challenges by systematically evaluating algorithm performance across comprehensive test instances~\cite{smithmiles2023instance}. It has been used to assess algorithms in fields as diverse as autonomous vehicle driving~\cite{Neelofar24features,crespo-rodriguez24autonomous}, the quadratic~\cite{FENNICH2024102} and multi-dimensional~\cite{SCHERER2024106747} knapsack problems, sports timetabling~\cite{VANBULCK2024575}, and multi-fidelity surrogate modelling~\cite{ANDRESTHIO2024104207}, and to generate new instances for the inventory routing problem~\cite{SKALNES2024992}, black-box continuous single-~\cite{Munoz20} and multi-objective~\cite{Yap22} optimisation, to mention a few, thanks to the accessibility of its MATLAB software tools~\cite{ISAMATLAB} and the cloud computing platform MATILDA (Melbourne Algorithm Test Instance Library with Data Analytics)~\cite{MATILDA}. MATILDA also provides a collection of ISA results and other meta-data for downloading for several well-studied problems from Optimisation and Machine Learning. However, MATILDA suffers from scalability issues due to its MATLAB-native computing code. This paper introduces \texttt{instancespace}, our implementation of ISA for Python, which replicates the functionality of the MATLAB code as described in~\cite{smithmiles2023instance}, while adding flexibility by better encapsulating the tasks within the ISA pipeline. Moreover, its installation through \texttt{pip} facilitates performing a local analysis and integrating with other benchmarking tools such as IOHexperimenter~\cite{IOHexperimenter}.

As a complete Python implementation of ISA, \texttt{instancespace} significantly enhances algorithm testing. First, providing more insightful and objective metrics for evaluation contributes to developing more reliable AI systems by supporting the creation of comprehensive test suites. Practically, \texttt{instancespace} allows users to process their algorithm performance data, visualise the instance space, and receive automated recommendations for selecting the best algorithm for new instances. While related work in the broader scientific community includes benchmark-based approaches to algorithm testing and algorithm selection methods such as meta-learning, \texttt{instancespace} offers a visual, data-driven, and instance-centric approach, filling a critical gap in existing AI validation and testing methods.

Second, the Python implementation allows \texttt{instancespace} to be integrated into existing systems without extensive modifications, making it more accessible for developers and researchers. Python's widespread adoption in academia and industry ensures that \texttt{instancespace} benefits from the ecosystem of libraries and tools, improving its functionality. Compared to MATLAB, Python's speed and scalability make it ideal for deploying services that require handling large-scale data or frequent analysis. The deployability of Python-based solutions allows users to integrate \texttt{instancespace} into pipelines with quicker iterations. Additionally, Python's familiarity among users reduces the learning curve, making \texttt{instancespace} easier to adopt and use effectively.

\section{Software Description}

\texttt{instancespace} implements the Instance Space Analysis (ISA) methodology~\cite{smithmiles2023instance}, facilitating the exploration and evaluation of algorithm performance across diverse test instances. To function, \texttt{instancespace} requires computational experimentation and the corresponding metadata generated from these experiments. The metadata contains essential information, such as the test instances' features and the algorithms' performance on these instances. Once provided, \texttt{instancespace} processes this metadata to construct an instance space, representing the test cases of the algorithm.

The instance space is a 2D projection that visualises the relationship between test instances and algorithm performance. \texttt{instancespace} analyses this space to make recommendations for untested instances and identify regions where algorithms are likely to perform well. These areas, referred to as "footprints," represent generalised regions where good (defined by the user) or best performance is inferred based on observed data.

Following the structure of the ISA methodology described in~\cite{smithmiles2023instance}, \texttt{instancespace} follows a six-stage process. Each stage addresses different aspects of instance space construction and analysis. Although the methodology follows a structured sequence, \texttt{instancespace} performs dependency resolution, ensuring that the stages can be run independently when required. This flexibility allows the software to process metadata efficiently and adapt to user needs. Moreover, \texttt{instancespace} requires a configuration file in JSON format to specify the options and parameters for each stage of the process.

\begin{description}
    \item[Preprocessing] organises and ensures that the metadata is formatted correctly. \texttt{instancespace} standardises feature values, handles missing data, and prepares the dataset for further analysis. Preprocessing also includes transformations, such as normalising feature values, to ensure that the data is suitable for dimensionality reduction and machine learning tasks.

    \item[PRELIM] prepares the metadata by defining a performance threshold that distinguishes between good and bad performance for each algorithm. It also bounds and normalises feature values, reducing the effect of outliers and ensuring that data is suitable for the next steps in the analysis. PRELIM applies transformations to the metadata and generates a binary measure indicating whether an algorithm performs well.

    \item[SIFTED] selects the most relevant instance features for analysis. It calculates the correlation between instance features and algorithm performance, identifying which features best explain the difficulty of test instances. By eliminating redundant or weakly correlated features, SIFTED ensures that only the most informative features are used in constructing the instance space.

    \item[PILOT] performs dimensional reduction, projecting the high-dimensional instance space onto two dimensions. This stage uses two approaches for dimensional reduction: the analytical approach, which uses eigenvalue projection, and the numerical approach, which minimises a high-dimensional curve using the BFGS algorithm. By reducing the dimensionality, PILOT enables the visualisation of the instance space and allows for the interpretation of the relationships between instance properties and algorithm performance.

    \item[CLOISTER] defines the boundaries of the instance space by projecting the theoretical or empirical limits of instance features. This step ensures that the instance space accurately reflects the range of possible test instances. By identifying regions where no instances exist, CLOISTER provides insights into the diversity and adequacy of the test instances and helps guide the generation of new instances to fill gaps in the instance space.

    \item[PYTHIA] automates algorithm selection using machine learning models trained on the instance space. PYTHIA predicts which algorithm will perform best based on the feature values of new, untested instances. The models are Support Vector Machines (SVMs), and their predictions are based on the performance of algorithms across the existing instance space.

    \item[TRACE] constructs algorithm footprints, representing regions in the instance space where an algorithm performs well. To refine these footprints, removing regions with weak or conflicting evidence about algorithm performance. The resulting footprints objectively measure an algorithm's strengths and weaknesses, offering insights into where it will most likely succeed.
\end{description}

\section{Software Architecture}

\texttt{instancespace} software architecture is built around a modular, stage-based design that ensures flexibility, scalability, and ease of maintenance. The software comprises discrete, interconnected components implementing the Instance Space Analysis (ISA) pipeline. Each component is designed as a separate class to perform a specific task, supporting customisation and extensibility.

At the core of \texttt{instancespace} is the \texttt{InstanceSpace} class, which orchestrates the interaction between the different stages of the pipeline. This class is the main controller, managing the data flow, ensuring process integrity, and handling external inputs such as experiment configurations and metadata. The \texttt{StageBuilder} is responsible for constructing the execution pipeline by connecting the stages and managing data dependencies. This allows dynamic addition or removal of stages, offering flexibility for specific research needs.

The architecture comprises a well-defined stage, each encapsulated within its class. Key classes are the \texttt{PreprocessingStage}, \texttt{PilotStage}, \texttt{PythiaStage}, \texttt{SiftedStage}, and \texttt{TraceStage}. These stages transform and analyse data progressively. Each stage performs specific functions: data preprocessing, dimensionality reduction, feature selection, algorithm footprint analysis, and performance-based evaluation. All stages adhere to a standard interface, enforcing consistent input/output behaviour via static typing and promoting uniformity across the system. 

Parallel execution is enabled through the \texttt{ParallelOptions} configuration, which ensures that large-scale experiments and computationally intensive tasks can leverage multi-core processing environments, improving scalability. The architecture also incorporates error handling, type safety, and extensibility through Python's type annotations and \texttt{NamedTuple} data structures, ensuring the system remains maintainable and adaptable for future enhancements.

\begin{figure}[!t]
    \centering
    \includegraphics[width=1.0\textwidth]{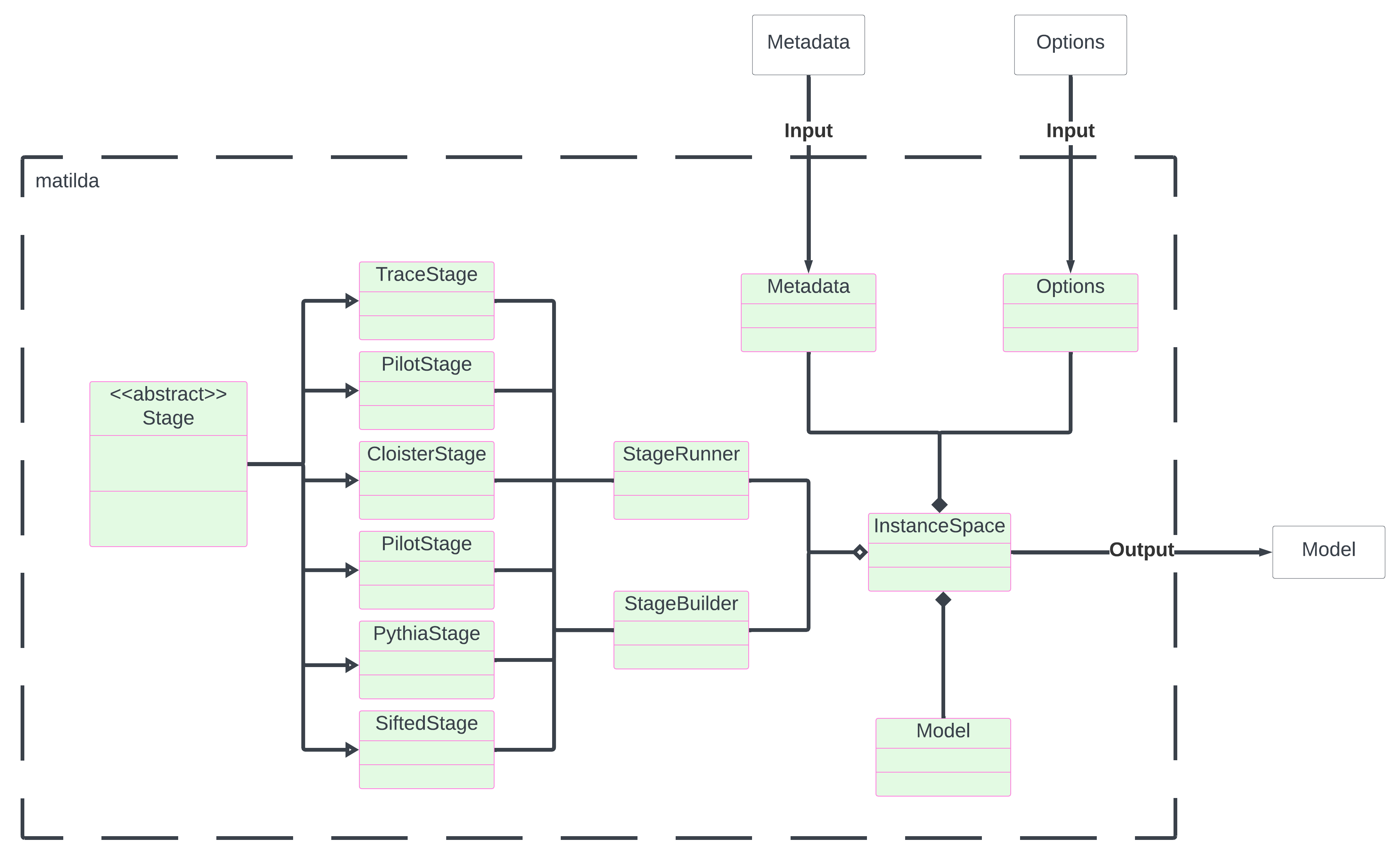}
    \caption{UML Diagram of \texttt{instancespace}}
    \label{fig:umldiagram}
\end{figure}

\subsection{Software Functionalities}

\texttt{instancespace} provides a set of functionalities designed to support a wide range of algorithm evaluation, selection, and instance space analysis tasks. 

\begin{description}
    \item[Instance Space Construction] \texttt{instancespace} allows users to construct an instance space from metadata and configuration files (e.g., \texttt{metadata.csv} and \texttt{options.json}), which maps the problem space where algorithms are tested. This enables the visualisation and exploration of algorithm performance across varying instances, helping users identify where algorithms perform well or poorly.
    
    \item[Stage-Based Execution] The modular architecture supports the execution of various stages, such as \texttt{PrelimStage}, \texttt{PilotStage}, \texttt{CloisterStage}, and \texttt{TraceStage}. Each stage represents a unique step in the analysis pipeline, transforming input data and computing metrics that feed into subsequent stages. Users can execute the entire pipeline or run individual stages for targeted analysis.
    
    \item[Dimensionality Reduction] Through techniques like Principal Component Analysis (PCA) and others, \texttt{instancespace} reduces complex, high-dimensional feature spaces to two-dimensional representations. This simplification facilitates intuitive visualisation of algorithm performance, aiding in interpretation and decision-making.
    
    \item[Algorithm Footprint Analysis] Using the \texttt{TraceStage}, \texttt{instancespace} generates geometric footprints representing regions of good, best, and beta performance for algorithms. These footprints provide a visual and statistical understanding of where algorithms excel and where they struggle in the instance space.
    
    \item[Automated Algorithm Selection] \texttt{instancespace} includes the \texttt{PythiaStage}, which utilises machine learning models such as Support Vector Machines (SVMs) to recommend the best-performing algorithms for untested instances. This functionality allows for efficient algorithm selection without exhaustive testing across all instances.
    
    \item[Synthetic Instance Generation] \texttt{instancespace} supports the generation of synthetic test instances to fill gaps in the instance space, ensuring a more comprehensive evaluation of algorithm performance. This is achieved through techniques like clustering and genetic algorithms.
    
    \item[Performance Metrics and Footprint Refinement] The \texttt{TraceStage} not only calculates geometric footprints but also refines them by removing contradictions and enhancing footprint purity. This allows users to evaluate key metrics like footprint area, density, and purity, providing a detailed performance summary for each algorithm.
\end{description}

\section{Illustrative Examples}

This section demonstrates how the \texttt{instancespace} framework operates using a real-world scenario. We guide the reader through the key stages of the Instance Space Analysis (ISA) process, showcasing how the software leverages metadata and options to analyse algorithm performance across various problem instances.

\subsection{Step 1: Loading Metadata and Options}

The first step in using \texttt{instancespace} involves loading the necessary configuration files: \texttt{metadata.csv}, which contains the problem instances and their features, and \texttt{options.json}, which configures the parameters for different stages in the analysis pipeline.

{\small
\begin{verbatim}
import sys
from pathlib import Path
import os
    
from matilda import \texttt{instancespace}
from matilda.data import metadata, options
    
# Define the paths for metadata and options
script_dir = Path(os.path.abspath('')) / "tests" / "test_data" / "demo"
metadata_path = script_dir / "metadata.csv"
options_path = script_dir / "options.json"
    
# Load metadata and options
metadata_object = metadata.from_csv_file(metadata_path)
options_object = options.from_json_file(options_path)
    
if metadata_object is None or options_object is None:
    print("ERR: File reading failed!")
    sys.exit()
\end{verbatim}}

This step initialises the system by reading metadata and options from their respective files. The \texttt{metadata.csv} file defines the instances, features, and algorithms used in the analysis, while \texttt{options.json} specifies parameters for configuring the various stages of the pipeline. The files are loaded using the \texttt{from\_csv\_file()} and \texttt{from\_json\_file()} functions, which parse the content into \texttt{Metadata} and \texttt{InstanceSpaceOptions} objects, respectively.

% \begin{itemize}
%     \item \textbf{Metadata:} The CSV file contains columns for instance labels, feature values, and algorithm performance metrics.
%     \item \textbf{Options:} The JSON file includes configuration options for parallelism, performance thresholds, dimensionality reduction, feature selection, and other settings.
% \end{itemize}

\subsection{Step 2: Constructing the Instance Space}

Once the metadata and options are loaded, the next step is constructing the instance space. This is done by initialising the \texttt{InstanceSpace} class, which connects the metadata and options and creates a pipeline of stages for the analysis.

{\small
\begin{verbatim}
from matilda.stages.cloister import CloisterStage
from matilda.stages.pilot import PilotStage
from matilda.stages.prelim import PrelimStage
from matilda.stages. preprocessing import PreprocessingStage
from matilda.stages.pythia import PythiaStage
from matilda.stages.sifted import SiftedStage
from matilda.stages.trace import TraceStage

# Construct the instance space and define the stages of the pipeline
instance_space = \texttt{instancespace}(
    metadata_object,
    options_object,
    stages=[
        PreprocessingStage,
        PrelimStage,
        SiftedStage,
        PilotStage,
        PythiaStage,
        CloisterStage,
        TraceStage,
    ],
)

# Print the order of stages for verification
print(instance_space._runner._stage_order)
\end{verbatim}}

The \texttt{InstanceSpace} object is initialised with the loaded metadata and options. A series of stages is defined, forming a pipeline that will be executed to process the instance space. The stages include:

\begin{itemize}
    \item \texttt{PreprocessingStage}: Cleans and normalizes the data.
    \item \texttt{PrelimStage}: Performs preliminary analysis and feature selection.
    \item \texttt{SiftedStage}: Applies feature correlation and clustering for dimensionality reduction.
    \item \texttt{PilotStage}: Tests algorithms on selected instances.
    \item \texttt{PythiaStage}: Recommends the best-performing algorithms.
    \item \texttt{CloisterStage}: Groups similar instances based on their performance.
    \item \texttt{TraceStage}: Generates performance footprints for algorithms.
\end{itemize}

This modular approach allows flexibility, where stages can be rearranged or skipped depending on the analysis requirements.

\subsection{Step 3: Running the Analysis Pipeline}

After defining the instance space and the stages, the pipeline calls the \texttt{build} method on the \texttt{InstanceSpace} object. This initiates the sequence of transformations, from data preprocessing to algorithm footprint generation.

{\small
\begin{verbatim}
# Execute the pipeline by building the instance space
instance_space.build()
\end{verbatim}}

The \texttt{build} method executes the stages defined in the pipeline sequentially. Each stage processes the data and passes it to the next stage, ultimately yielding an analysis of algorithm performance across the instance space. This includes reducing the dimensionality of the data, selecting relevant features, clustering instances, and generating performance metrics.

\subsection{Step 4: Analysing the Results}

After the pipeline has run, the analysis results can be accessed for further interpretation. We retrieve and print the processed data matrix from the \texttt{InstanceSpace} model in this example. This matrix contains the transformed feature vectors for each problem instance.

{\small
\begin{verbatim}
# Access the model and retrieve the processed data matrix
model = instance_space.model
data_matrix = model.data.x
print(data_matrix)
\end{verbatim}}

The \texttt{model} object in \texttt{InstanceSpace} holds the analysis results, including the transformed instance space. The \texttt{data\_matrix} variable contains the feature vectors for each problem instance after the data has passed through all pipeline stages. These vectors can be used for visualisation or further analysis.

{\small
\begin{verbatim}
[[ 0.23, -1.27,  0.56, ...],
 [ 1.01,  0.75, -0.63, ...],
 ... ]
\end{verbatim}}

The output shows a matrix where each row corresponds to an instance, and each column corresponds to a transformed feature. This matrix is the final representation of the instance space, reduced in dimensionality and optimised for algorithm performance evaluation.

\subsection{Step 5: Algorithm Footprint Analysis}

One of the key functionalities of \texttt{instancespace} is its ability to generate geometric footprints for algorithms based on their performance across the instance space. In this step, we demonstrate how the \texttt{TraceStage} generates performance footprints using clustering techniques.

{\small
\begin{verbatim}
# Execute the TraceStage to analyse algorithm footprints
footprints = instance_space.model.trace_stage_results
print(footprints.summary)
\end{verbatim}}

The \texttt{TraceStage} analyses the regions in the instance space where algorithms perform well (or poorly) by generating geometric footprints. These footprints are clustered using techniques such as DBSCAN to group similar instances based on their performance. The \texttt{summary} output provides a detailed breakdown of each footprint's area, density, and purity, helping users understand the strengths and weaknesses of the algorithms in different regions of the instance space.

\begin{table}[!t]
\centering
\caption{Footprint Analysis}
\footnotesize
\begin{tabular}{lp{1.2cm}p{1.2cm}p{1.2cm}p{1.2cm}p{1.2cm}p{1.2cm}}
\toprule
Row & Area Good Normalized & Density Good Normalized & Purity Good & Area Best Normalized & Density Best Normalized & Purity Best \\ \midrule
FFD     & 1.034 & 0.962 & 0.765 & 0 & 0 & 0 \\ \midrule
DJD1\_2 & 1.005 & 0.976 & 0.86  & 0 & 0 & 0 \\ \midrule
\end{tabular}%
\end{table}

\begin{table}[!t]
\centering
\footnotesize
\caption{Average Performance Metric With SVM Accuracy and Parameters}
\begin{tabular}{lp{2.2cm}p{2.2cm}p{2.2cm}p{2.2cm}}
\toprule
Algorithm & Avg. Performance (Std. Performance) & Actual Percentage of "Good" Performances & Avg. Performance (Selected Instances) & SVM Accuracy and Parameters (Overall Accuracy (\%), Precision, Recall) \\ \midrule
FFD       & 3.43 (6.242) & 0.815 & 3.563 (6.669) & 84.3, 88.2, 93.2 \\ \midrule
DJD1\_2   & 3.531 (6.201) & 0.79  & 3.846 (6.766) & 89.1, 91.4, 95.2 \\ \midrule
Oracle    & 3.173 (6.262) & 1     & ---           & ---              \\ \midrule
Selector  & 3.27 (6.263)  & 0.929 & 3.288 (6.284) & --, 92.9, ---    \\ \bottomrule
\end{tabular}%
\end{table}

\begin{figure}[!t]
    \centering
    \includegraphics[width=0.70\linewidth]{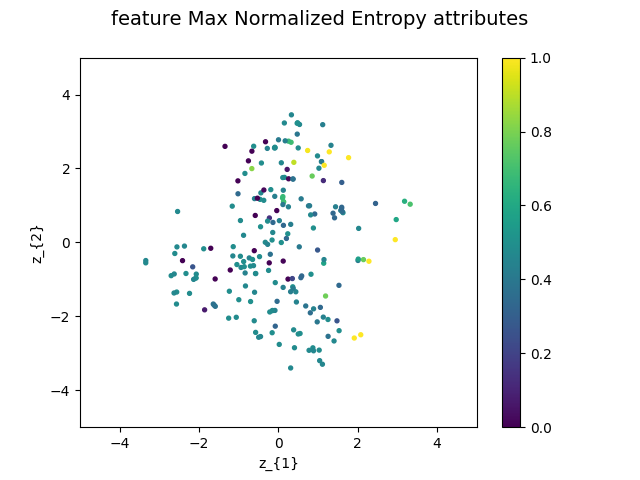}
    \caption{An example of a plot produced by \texttt{instancespace}, where each dot represents an instance, and the color gradient represents the feature value.}
    \label{fig:ExamplePlot}
\end{figure}

\section{Impact}

Instance Space Analysis (ISA) provides a methodology for evaluating algorithms beyond traditional benchmarking methods. By visualising the instance space and the relationship between instance properties and algorithm performance, researchers can systematically explore how various instance characteristics affect algorithm behavior, leading to more informed algorithm design and selection processes. The synthetic generation of test instances based on gaps in the instance space also supports new investigations into algorithm robustness, particularly in previously underexplored areas.

\texttt{instancespace} helps existing research questions related to algorithm evaluation. By providing unbiased metrics for algorithm performance, \texttt{instancespace} allows for more detailed performance assessments than standard average-based benchmarking. Researchers can now explore how algorithms perform across various test instances, visualise their strengths and weaknesses, and identify regions where specific algorithms excel or fail. This level of insight is crucial for developing and testing algorithms in fields such as optimisation, machine learning, and scheduling, where \texttt{instancespace} has already been applied successfully.

In practice, \texttt{instancespace} can help researchers conduct algorithm testing. Its automated algorithm selection and instance generation tools streamline algorithm benchmarking, reducing the time and effort required to evaluate new algorithms. Additionally, by enabling users to upload their algorithms and benchmark them against existing instance spaces, \texttt{instancespace} promotes a more collaborative and transparent approach to algorithm testing.

% \texttt{instancespace} is widely used within its intended user group, with over 150 researchers signing up for applications in various problem domains, including combinatorial optimisation, machine learning, and time series forecasting. The tool has been employed in numerous research publications, demonstrating its utility in diverse fields. The growing repository of problem classes in \texttt{instancespace}, which includes optimisation problems such as graph coloring, the traveling salesman problem, and mixed-integer programming, reflects its broad applicability. These contributions are citable, and researchers are encouraged to add their problem instances to further expand the platform's utility.

While \texttt{instancespace}'s use is primarily academic, its applications extend beyond the research community. The tool's ability to objectively assess algorithmic performance and suggest the best algorithm for a given instance has potential value in commercial settings, particularly in industries with prevalent optimisation problems. % Although no spin-off companies have emerged from this work thus far, the possibility exists for \texttt{instancespace}'s methodology to influence future commercial algorithm development, particularly in sectors such as logistics, finance, and automated scheduling.

\section{Conclusions}

\texttt{instancespace} represents a significant development for testing and evaluating algorithms by applying Instance Space Analysis (ISA). By providing a visual and data-driven methodology, \texttt{instancespace} enables researchers to explore algorithm performance across a wide range of test instances and identify areas of strength and weakness. The ability to generate synthetic test instances, combined with automated algorithm selection, positions \texttt{instancespace} as a powerful tool for enhancing the robustness of AI models, particularly in safety-critical applications.

The package flexibility ensures adaptability to various problem classes, from optimisation to machine learning. \texttt{instancespace} enhances the scientific process by offering objective, unbiased metrics and streamlining the testing process, making it accessible to a broader audience.

Future work will focus on expanding the range of problems supported by \texttt{instancespace} and refining its algorithm selection capabilities. As AI systems continue to be integrated into increasingly complex and high-stakes environments, the need for robust, comprehensive testing frameworks like \texttt{instancespace} will only grow in importance.

\section*{Acknowledgements}

Most of the code base was developed as part of the subject SWEN90017-18, using as a basis the original ISA MATLAB code developed by Mario Andrés Muñoz, by students Yusuf Berdan Guzel, Kushagra Khare, Dong Hyeog Jang, Kian Dsouza, Nathan Harvey, Junheng Chen,  Tao Yu, Xin Xiang, Jiaying Yi, and Cheng Ze Lam. Mario Andrés Muñoz and Ben Golding mentored the team, and Mansooreh Zahedi coordinated the subject. The Australian Research Council provided partial funding for developing this code through grant IC200100009.

\bibliographystyle{elsarticle-num} 
\bibliography{references}

\end{document}